\documentclass[12pt,preprint]{aastex}

\slugcomment{to appear in ApJL}

\shorttitle{Ly$\alpha$ deficiency of luminous LBGs at $z\sim5-6$}
\shortauthors{Ando et al.}

\begin{document}


\title{Deficiency of large equivalent width Ly$\alpha$ emission 
in luminous Lyman break galaxies at $z\sim5-6$?\altaffilmark{1}}


\author{Masataka Ando\altaffilmark{2}, Kouji Ohta\altaffilmark{2}, Ikuru Iwata,\altaffilmark{3}}
\author{Masayuki Akiyama\altaffilmark{4}, Kentaro Aoki\altaffilmark{4},
 and Naoyuki Tamura,\altaffilmark{4,5}}

\altaffiltext{1}{Based on data collected at Subaru Telescope, which is operated
by the National Astronomical Observatory of Japan.}
\altaffiltext{2}{Department of Astronomy, Kyoto University, Kyoto
606-8502, Japan.; andoh@kusastro.kyoto-u.ac.jp, ohta@kusastro.kyoto-u.ac.jp}
\altaffiltext{3}{Okayama Astrophysical Observatory, National
Astronomical Observatory, Kamogata, Okayama 719-0232, Japan.; iwata@oao.nao.ac.jp}
\altaffiltext{4}{Subaru Telescope, National Astronomical Observatory of
Japan, 650 North A'ohoku Place, Hilo, HI 96720.;
akiyama@subaru.naoj.org, kaoki@subaru.naoj.org, naoyuki@subaru.naoj.org}
\altaffiltext{5}{Department of Physics, University of Durham, Durham DH1 3LE, UK.}

\begin{abstract}
We report a deficiency of luminous Lyman break galaxies (LBGs)
with a large rest-frame equivalent width (EW$_{\rm rest}$)
of Ly$\alpha$ emission at $z\sim5-6$.
Combining our spectroscopic sample of LBGs at $z\sim5$ and those from
the literature, we found that luminous 
LBGs at $z\sim5-6$ generally show weak Ly$\alpha$ emissions,
while faint LBGs show a wide range of Ly$\alpha$ EW$_{\rm rest}$
 and tend to have 
strong (EW$_{\rm rest}\gtrsim20$\AA) Ly$\alpha$ emissions;
i.e., there is a deficiency of strong Ly$\alpha$ emission in
luminous LBGs.
There seems to be a threshold UV luminosity for the deficiency;
it is $M_{1400}= -21.5 \sim -21.0$ mag, which is close to or somewhat
 brighter than the $M_*$ of 
the UV luminosity function at $z\sim5$ and 6.
Since the large EW$_{\rm rest}$ of Ly$\alpha$ emission can be seen among
the faint LBGs, the fraction of Ly$\alpha$ emitters in LBGs
may change rather abruptly with the UV luminosity.
If the weakness of Ly$\alpha$ emission is due to dust absorption,
the deficiency suggests that luminous LBGs at $z=5-6$ tend to
 be in dusty and more chemically evolved environments
and that they start star formation earlier than faint LBGs,
though other causes cannot be ruled out. 
\end{abstract}





\section{Introduction}
Recently, a number of Lyman break galaxies (LBGs) at $z\sim5-6$
have been found, and 
their photometric properties and spatial distribution have been studied
extensively.
Studies have been made on a UV luminosity
function (UVLF) as well as a star formation rate (SFR) 
density at the epoch \citep[e.g.,][]
{Iwa03,Bun04,Yan04,Dick04,Ouchi04a,Shima05,Bou06}
and on  a two-point correlation function which constrains the
mass of the dark halo where a LBG (or LBGs) resides  \citep[e.g.,][]
{Ouchi04b,kashi06}.
Furthermore, analysis of spectral energy distributions (SEDs) has 
constrained their stellar masses, ages, degree of extinction, and so on
for the $z\gtrsim5$ LBGs \citep[e.g.,][]{Eyl05,Yan05,Chary05,Schaerer05}.

However, the spectroscopic properties of LBGs at $z\gtrsim5$ are still unknown.
Only about a dozen spectra have been published so far
\citep[][]{Spi98,Dey98,Wey98,Daw02,Lehn03,Sta03,Sta04,Nagao04,Dow05,Nagao05},
and extensive follow-up spectroscopies have been currently made.
Among these, most of the spectra show only Ly$\alpha$ emission,
and virtually no information can be drawn from their continuum spectrum.
\citet{frye} successfully detected the continuum and line features of
gravitationally lensed
galaxies at $z=4\sim5$ and have found the presence of interstellar absorption 
lines.
More recently, \citet{Dow05} detected the continuum 
and interstellar absorption lines of an gravitationally lensed
LBG at $z=5.5$ with a good signal-to-noise (S/N) ratio
with an extremely long (22.3 hours) exposure time.

We are also conducting a spectroscopic study of LBGs at $z\sim5$
found by \citet{Iwa03} and Iwata et al. (in preparation), 
and \citet{Ando04} present the first results.
The obtained spectra show no or weak Ly$\alpha$ emission in contrast to 
LBGs at $z\sim3$ \citep[e.g.,][]{Shap03,Iwa05a},
and rather strong low-ionization interstellar (LIS) absorption lines of
which equivalent widths (EWs) are comparable to those seen in LBGs at
$z\sim3$ \citep{Shap03}.
We also found 
the Ly$\alpha$ emissions are redshifted by about $500-700$ 
km s$^{-1}$ relative to the interstellar absorption lines in a part of
them, which suggests the presence of an outflow in the LBGs at $z\sim5$.

In this letter, we report a possible UV luminosity 
dependence of Ly$\alpha$ emission of LBGs at $z\sim5-6$ using our previous 
and new spectroscopic sample 
together with those from the literature, and we discuss some possible causes.
Throughout this paper, we adopt flat $\Lambda$ cosmology, $\Omega_M=0.3$,
$\Omega_{\Lambda}=0.7$, and $H_0=70$ km s$^{-1}$Mpc$^{-1}$. 
The magnitude system is based on AB magnitude \citep{Oke83}.

\section{Spectroscopic Samples of LBGs at $z\sim5$ and $z\sim6$}
The spectroscopic sample of LBGs at $z\sim5$ is taken mainly
from \citet{Ando04} and our recent results taken with FOCAS
\citep{kashi} attached to the Subaru telescope \citep{Iye04}.
The former sample is obtained around the GOODS-N field
and the latter one is obtained around the J0053+1234 field
\citep[see][]{Ando05}. 
The LBG selection of our sample is based on $V-I_C$ and $I_C-z'$ colors,
and details are described in 
\citet{Iwa03,Iwa05b} and Iwata et al. (in preparation).
Magnitudes of the spectroscopic sample in the $I_C$ band are
$24.0-25.3$, and the redshift coverage is $z=4.3-5.2$.
The spectroscopic observations and data reductions are achieved by
almost the same way for both samples as described in \citet{Ando04}.
For the sample in the GOODS-N field, resulting spectra of 7 bright 
($I_C\leq25.0$) LBGs are shown in Figure 1 of \citet{Ando04}.
They show the continuum depression shortward of redshifted
Ly$\alpha$ and some LIS metal absorption lines,
indicating secure identifications of the redshifts.
For the sample in the J0053+1234 field,
2 bright ($I_C\leq25.0$) LBGs were confirmed with the similar
features to the GOODS-N sample,
and 2 faint ones were identified with a strong
(EW$_{\rm rest}\sim40-80$\AA ) Ly$\alpha$ emission line.
The average Ly$\alpha$ EW$_{\rm rest}$ of our 9 bright spectroscopic
LBGs is $\sim6\pm7$\AA , and that of 2 faint ones is $\sim60\pm20$\AA.
In addition to our sample, we take the spectral data of similar redshifts
from the literature \citep{Spi98,Dey98,Daw02,Lehn03}.
The number in the sample is 6, and the redshift coverage is $z=4.8-5.3$.
Since the sample of \citet{frye} is gravitationally lensed and their
magnitude correction contains a relatively large uncertainty (maximally
1.5 mag),
it is excluded in the following discussion.

We also compiled spectroscopic results of LBGs at $z\sim6$ 
\citep{Wey98,Lehn03,Sta03,Sta04,Nagao04,Dow05,Nagao05}.
The number of galaxies in the sample is 9,
and the redshift coverage is $z=5.5-6.3$.
The object by \citet{Dow05} is a gravitationally lensed galaxy.
But its amplification effect ($\sim0.3$ mag at most) is small,
and thus we include it.
Note that the objects by \citet{Nagao04,Nagao05} are the $i'$-dropout
objects with narrow band (NB921) depression;
a strong Ly$\alpha$ emission largely contributes to the $z'$-band light.

\section{Result and Discussion}

\subsection{Deficiency of UV luminous LBGs with a large Ly$\alpha$ equivalent width}

With the samples described in section 2, 
we adopted EWs of Ly$\alpha$ emission appeared in each paper.
For the objects of our spectroscopic sample of LBGs at $z\sim5$ 
in the J0053+1234 field,
we measured EWs by the same way used in \citet{Ando04}.
The uncertainties of the EWs are estimated to the 30-50\%.
For the sample of \citet{Lehn03}, we measured the EWs from their
Figure 4 because the EWs of individual objects were not presented.
The uncertainties of EWs are roughly 50\%.
For all these values of EWs, the absorption for the emission line 
by intergalactic matter (IGM)
was not corrected, and the absorption component of
Ly$\alpha$ was not included.
Figure 1 shows rest EWs of Ly$\alpha$ emissions plotted against the
rest-frame UV absolute magnitudes.
Filled circles show our spectroscopic results, and 
filled squares refer to the results from the literature.
Open squares represent those of LBGs at $z\sim6$.
We converted the observed broad band magnitude to the rest-frame 1400\AA\
magnitude assuming a continuum slope $\beta$
($f_{\lambda}\propto\lambda^{\beta}$) of $-1$ 
which is a typical value for LBGs at $z\sim3$ \citep{Shap03}. 
The effect of the uncertainty of the slope is small and estimated to be 
typically $0.1-0.2$ mag.
For the objects whose adopted broad band contained the wavelength
region shortward of Ly$\alpha$,
we corrected for the contributions by Ly$\alpha$ emission and IGM 
absorption \citep{Madau95} to derive the UV absolute magnitudes.
From the UV absolute magnitude using the relation by \citet{Madau98}, 
we also show the SFR estimated at the upper abscissa.

As seen in Figure 1, there are no UV luminous
($M_{1400}\lesssim-21.0$ mag) LBGs at $z\sim5$ with
strong (EW$_{\rm rest}\gtrsim$20\AA )
Ly$\alpha$ emission, while UV faint
ones show wide range of EW$_{\rm rest}$ and tend to have stronger Ly$\alpha$
emission than UV luminous LBGs on average.
In addition, there seems to be a UV magnitude threshold for LBGs with
strong Ly$\alpha$ emission around
$M_{1400}\sim-21.0$ mag which is almost the same as the $M_{\ast}$
magnitude of the UVLF of our $z\sim5$ LBG sample \citep{Iwa03} and that
of \citet{Ouchi04a}.
This trend still holds if the data by \citet{frye} are considered.

A similar deficiency of the
luminous LBGs with strong Ly$\alpha$ emission seems to hold at $z\sim6$,
although the sample size is quite small especially for a luminous part
($M_{1400}\lesssim-21.5$ mag).
There seems to be the threshold magnitude around $M_{1400}\sim-21.5$
mag which is close to the $M_{\ast}$ of the UVLF at $z\sim6$ of
\citet{Bun04} and $\sim 1.4$ mag brighter than that of \citet{Bou06}.
The threshold magnitude is $\sim0.5$ mag brighter
than that of LBGs at $z\sim5$, which might suggest its evolution.
But the current sample number is not large enough to be 
definitively to illustrate the evolution of the threshold magnitude.

We note that the deficiency of the
luminous LBGs with strong Ly$\alpha$ emission is not due to observational
bias at least for our spectroscopic sample.
First, the minimum detectable EW$_{\rm rest}$ of the Ly$\alpha$ emission
in our spectroscopic survey is $\sim$10\AA\ for luminous  
(corresponding to $I_C\leq25.0$)
LBGs at $z\sim5$, and $\sim$30\AA\ for faint 
ones, respectively, in the wavelength regions where
night sky emission lines are not strong.
Therefore, we should detect large EW Ly$\alpha$ emission lines among
luminous LBGs, if there are such objects.
Second, the number of the {\it observed} luminous LBGs 
in our present spectroscopic sample (22 objects) 
is larger than the number of the faint ones (12 objects).
Thus, if the fraction of objects with strong Ly$\alpha$ emission is the same
for UV luminous and faint LBGs, we should detect at least several
luminous LBGs with strong Ly$\alpha$ emission, unless we happened to 
choose LBG candidates with a small EW$_{\rm rest}$ in luminous
candidates selectively.
The Kolmogorov-Smirnov test would not be useful because
faint LBGs with a small 
EW$_{\rm rest}$ are expected to be missed in the present sample.

For the $i-z$ selected LBGs at $z\sim6$, there may be a selection bias
which leads to the apparent deficiency of LBGs with strong Ly$\alpha$
emission because the strong Ly$\alpha$ emission contributes largely to
the $i$-band flux and reduce the value of the $i-z$ color.
Expected $i-z$ colors of the LBGs with rest EW of 20(100)\AA\ are 
$\sim0.2$ mag ($\sim0.7$ mag) bluer
than $i-z=1.5$ mag, the color criterion to pick up $i$-dropout
objects, in the redshift range between $\sim5.9$ to $\sim6.0$ 
(between $\sim5.9$ to $\sim6.1$)
\footnote{The expected color and affected redshift range 
depend somewhat on the adopted spectral
templates of star-forming galaxies.}.
Among the samples we used here, \citet{Sta03} and \citet{Dow05}
utilized the $i-z$ selection method,
and the samples may suffer from the selection bias.
However, this bias should work independently of the UV luminosity,
thus it may not seriously affect the deficiency of luminous LBGs 
with strong Ly$\alpha$.
Note that the sample of LBGs at $z\sim5$ does not suffer from 
this selection bias.

Since the EW$_{\rm rest}$ increases with decreasing UV continuum luminosity
for a constant Ly$\alpha$ luminosity, the deficiency may just reflect 
the distribution of a constant Ly$\alpha$ luminosity.
In Figure 1, we show locations of constant Ly$\alpha$ luminosity 
corresponding to 5$\times$10$^{43}$, 2$\times$10$^{43}$, 10$^{43}$,
5$\times$10$^{42}$, and 10$^{42}$ erg s$^{-1}$
with dotted lines from top-left to bottom-right.
All the luminous LBGs with small EWs have Ly$\alpha$ luminosity
smaller than or equal to $10^{43}$ erg s$^{-1}$,
while about a half of the faint LBGs show Ly$\alpha$ luminosity
larger than $10^{43}$ erg s$^{-1}$.
Although the contrast is not so strong as compared with that 
seen in the EWs, again
there are no UV luminous LBGs having a large Ly$\alpha$ luminosity.
In any cases, the present sample is a combination of our 
spectroscopic data  and  those from the literature, 
and the sample size is small.
A more homogeneous and larger spectroscopic sample
is needed to examine whether this trend is definitive or not, 
although such data sets are hard to obtain even with currently available
8-10m telescopes.

\subsection{Lyman $\alpha$ emitters at $z\sim6$}
Since Ly$\alpha$ emitters (LAEs) are expected to have large 
rest EWs due to their selection method,
LAEs may be located in the upper left part
of the Figure 1.
We plot LAEs at $z\sim5.7$ and $z\sim6.6$ detected from narrow-band 
imaging data in Figure 1; 
crosses and pluses represent LAEs at $z\sim5.7$ \citep{Aji03} and
LAEs at $z\sim6.6$ \citep{Tani05}, respectively.
We adopted values of EWs from spectroscopic results for a part of LAEs at
$z\sim6.6$  \citep[triangles:][]{Tani05}.
The values of Ly$\alpha$ rest EWs are not corrected for IGM
absorption for the emission.
Most of the LAEs are UV faint objects and the
rest EWs distribution is similar to that of faint LBGs.
For the UV luminous LAEs, Ly$\alpha$ rest EWs are relatively
smaller than those of faint LAEs and faint LBGs, and again we can see
the deficiency of the strong Ly$\alpha$ emission in UV luminous LAEs.
A recent survey of LAEs at $z=5.7$ in the Subaru Deep Field
\citep{Shima06} also shows the same tendency (see their Figure 16).

These results imply that a fraction of LAEs in LBGs changes
with the UV luminosity at $z \sim 5-6$;
among UV luminous LBGs, there are only a few LAEs, while among faint LBGs
there are many LAEs with large EW of Ly$\alpha$ emission.
This fits the trend that the ratio of LAEs to LBGs at $z\sim5$ 
decreases with increasing UV luminosity \citep{Ouchi03},
although our results show a rather abrupt decrease of LAEs among luminous
LBGs. 
Thus, the LAEs are presumably a subset of faint LBGs, and are
not the luminous LBGs at the redshift.

\subsection{Possible origins of the deficiency}
Although we need further studies to confirm this 
deficiency of the luminous LBGs with a large Ly$\alpha$ EW$_{\rm rest}$,
we try to find possible causes for the trend and its
implications.
One possible cause is the systematic difference of the dust extinction
between luminous and faint LBGs.
Since there are significant correlations between gas metallicity, dust
extinction, and strength of LIS absorption lines in local star-forming
galaxies \citep[e.g.,][]{Hek98}, 
the presence of strong LIS absorption lines in the luminous LBGs of our
spectroscopic sample support this possibility.
If we assume the local relation between the EW of LIS absorptions 
and metallicity by \citet{Hek98}, an estimated gas metallicity for
our spectroscopic sample of luminous LBGs at $z\sim5$ is
12$+$log(O/H)$\sim8.0$ \citep[$\sim$1/5 solar: using the solar value 
from][]{Pri01}.
At this gas metallicity, the Ly$\alpha$/H$\beta$ ratio is reduced by
a factor of about 30 from the Case B assumption for local star-forming
galaxies \citep{Har88}.
If the luminous LBGs at $z\sim5$ are more chemically evolved 
than the faint LBGs, it is suggested that the luminous LBGs 
at $z\sim5$ started star formation relatively earlier than faint ones.
The clustering analysis of LBGs at $z\sim5$ shows that luminous
LBGs have a larger correlation length than faint ones, suggesting
luminous LBGs reside in a more massive dark halo
(e.g., \citet{Ouchi04b}; Iwata et al. in preparation).
The suggestions seem to fit the biased star formation scenario in 
the early universe; UV luminous LBGs at $z\sim5$ are in a more massive 
dark halo and have experienced star formation earlier than faint ones
residing in a less massive dark halo.
We do not find the significant relation between 
the $I_C-z'$ color and the EW$_{\rm rest}$ of Ly$\alpha$ emission
for our spectroscopic sample.
However, this is probably because the baseline separation between $I_C$
and $z'$ is too small to derive reliable $E(B-V)$ values under the
current photometric uncertainty.
In addition, the S/Ns of our spectra are too low and the wavelength coverage
is too small to reliably estimate the $E(B-V)$ values from their continuum.

The amount of HI gas in and surrounding the galaxy can affect the 
Ly$\alpha$ EWs.
If luminous LBGs have much more HI gas than faint ones,
it is possible that Ly$\alpha$ emission is selectively extinguished in
luminous LBGs, resulting in small EWs.
The presence of a large amount of HI gas in luminous LBGs
seems to fit the biased galaxy formation
scenario described above; luminous LBGs reside in a more massive dark
halo are expected to have more HI gas than faint ones in a less massive 
dark halo.

The age of a galaxy can also affect the EW of Ly$\alpha$.
A large Ly$\alpha$ EW$_{\rm rest}$ (100-200\AA ) is expected for very young
($<10-100$Myr) galaxies \citep[e.g.,][]{Cha93}.
Thus the luminous LBGs may be rather older than faint ones.
However, it is hard to claim so, if the dust exists in a galaxy.
In fact, using the results of SED fitting of LBGs at $z\sim3$, \citet{Shap01}
found that the composite spectrum of young (age $\leq10$Myr) LBGs
shows a small Ly$\alpha$ EW, while that of old (age $\geq1$Gyr) LBGs 
shows a large one.

Another possibility is velocity structure of the HI gas in/around the galaxy.
From Ly$\alpha$ imaging and spectroscopy of nearby star-forming
galaxies, EWs of Ly$\alpha$ do not necessarily depend on the 
gas metallicity. 
For example, \citet{kunth98,kunth}
claims that the kinematical property of the gas is
a dominant regulator of the Ly$\alpha$ escape probability;
galaxies with outflowing neutral gas tend to have large Ly$\alpha$ EWs,
while galaxies with static neutral gas tend to have small Ly$\alpha$ EWs.
\citet{Shap03} also pointed out the importance of the kinematical
feature for the LBGs at $z\sim3$, but their sense of effect on Ly$\alpha$ EW 
is opposite to that of \citet{kunth98}.
They found that LBGs with smaller Ly$\alpha$ EWs tend to have stronger LIS
absorptions and large velocity offsets of Ly$\alpha$ emission relative
to LIS absorption lines.
These facts can be explained
if the LBGs with smaller EWs have outflowing neutral 
gas with a larger velocity dispersion, because
the gas causes a broader Ly$\alpha$ absorption for
Ly$\alpha$ emission that results in a smaller EW of Ly$\alpha$,
a more redshifted (asymmetric) Ly$\alpha$ peak that is seen as the
larger velocity offset between Ly$\alpha$ and LIS absorption,
and stronger LIS absorption lines \citep{Shap03}.
We found the asymmetry of the Ly$\alpha$ emission line and the velocity
offset of Ly$\alpha$ emission relative to LIS
absorption lines in a part of our luminous spectroscopic sample, 
which implies the presence of the large-scale motion of
the neutral gas of LBGs at $z\sim5$ as well as at $z\sim3$.
Thus there is a possibility that the gas kinematics
affects the EW and the shape of profile of Ly$\alpha$ emission.

\acknowledgments

We are grateful to the FOCAS team, especially Youichi Ohyama, 
and the staff of the Subaru telescope for their support during our
observation.
We appreciate the anonymous referee for the comments which improved
this paper.
The preparations of the observation were in part carried out on 
the "sb" computer system
operated by the Astronomical Data Analysis Center (ADAC) and Subaru
Telescope of the National Astronomical Observatory of Japan.
MAs are supported by a Research Fellowship of the Japan Society for the
Promotion of Science for Young Scientists.
KO is supported by a Grant-in-Aid for Scientific Research from
Japan Society for the Promotion of Science (17540216).
This work is supported by the Grant-in-Aid for the 21st Century COE
"Center for Diversity and Universality in Physics" from the Ministry of
Education, Culture, Sports, Science and Technology (MEXT) of Japan.

\clearpage

\begin{figure}
\plotone{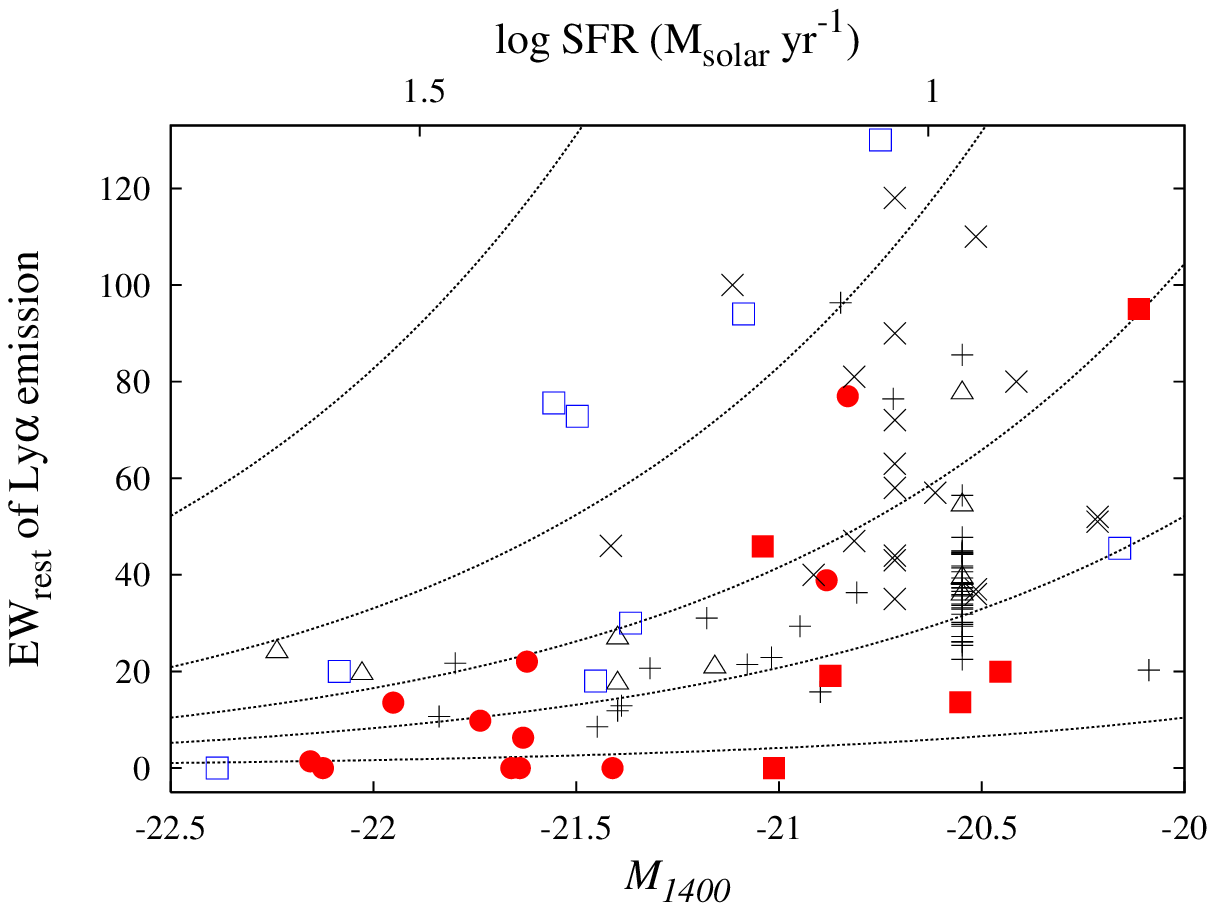}
\caption{
Rest-frame EWs of Ly$\alpha$ emission vs. absolute
magnitude at rest-frame 1400\AA\ for galaxies at $z\sim5$.
Filled circles show our previous \citep{Ando04} and newly added results.
Filled squares show data taken from the literature
\citep{Lehn03, Spi98, Daw02, Dey98}.
Open squares show the data of galaxies at $z\sim6$ taken from 
\citet{Wey98,Lehn03,Sta03,Sta04,Nagao04,Dow05,Nagao05}.
Crosses and pluses represent Ly$\alpha$ emitters at $z=5.7$ and $z=6.6$, 
respectively \citep{Aji03,Tani05},
obtained from narrow-band imaging.
A part of the sample of \citet{Tani05} have spectroscopic results of
 EW$_{\rm rest}$, and we plot them as triangles.
Star formation rate (SFR) estimated from UV absolute magnitude
 \citep{Madau98} is also shown at the top.
Dotted lines show Ly$\alpha$ EW$_{\rm rest}$ as a function of UV absolute
magnitude for a constant Ly$\alpha$ luminosity of 5$\times$10$^{43}$,
2$\times$10$^{43}$, 10$^{43}$, 5$\times$10$^{42}$, and 10$^{42}$ erg s$^{-1}$
from top-left to bottom-right.
}
\end{figure}

\end{document}